# Air-plasma-based all-optical temporal integration of broadband terahertz pulses

*Jia-Yu Zhao*

Terahertz Technology Innovation Research Institute, Terahertz Spectrum and Imaging Technology Cooperative Innovation Center, Shanghai Key Lab of Modern Optical System, University of Shanghai for Science and Technology, Shanghai 200093, China

E-mail: zhaojiayu@usst.edu.cn

**Abstract**

Platforms for all-optical computing possess photonic circuits made of silicon-based channels, metal wires or optical fibers, etc. However, such solid waveguides suffer from the lack of reversibility and reconfigurability if facing the next generation of light-guiding-light (LGL) computing scheme, which envisions circuitry-free and rapidly reconfigurable systems powered by dynamic interactions between light beams. Here, we proposed the ubiquitous air as a restorable LGL signal manipulation medium with transient air-plasma waveguide circuits. Briefly, by focusing femtosecond laser beams in the free space, the created atmospherical plasma filament array via photoionization was able to guide terahertz (THz) pulses along its epsilon-near-zero (ENZ) zone with a $1/f$-profile spectral response. Consequently, this achieved a time-domain integration of the THz pulse in broad bandwidth. When the pumping laser was sequentially turned off and on, this multi-filament air-plasma structure was erased and rebuilt within nano- and femto-seconds, respectively, allowing rapid and repeated rearrangements of the all-optical stage. Furthermore, this air-based LGL information processing approach is promising to pave the way towards all-optical calculations during free-space directional transmission of THz waves, by which means the delivered THz signal can be remotely controlled.

## 1. Introduction

All-optical computations, which overcome the speed limitation imposed by photoelectric-based systems, are done solely with light beams guided in circuits of silicon-substrate channels,[1,2] metal wires[3,4] or optical fibers,[5] etc. However, such hard waveguide circuitry suffers from the lack of reversibility and reconfigurability before stepping into the next generation of light-guiding-light (LGL) scheme, which envisions circuitry-free, rapidly reconfigurable and multilayered computing setups powered by dynamic interactions between light beams.[6] Thus recently, reversible soft hosts for LGL signal processing, e.g., photoresponsive gels[6] and photopolymer,[7] have been proposed in order to remove those rigid circuital components, attracting intense attention in the community.

Besides the above solid platforms,[1-7] the gaseous substance (e.g., air) can also be a potential candidate for the LGL material, which features the unique free-space property. For instance, when focusing femtosecond laser beams in air, the atmospherical plasma filaments array is created via photo-ionization,[8,9] which can further guide the propagation of terahertz (THz) waves,[10] hence realizing the LGL effect. Moreover, this gas-based stage can be rapidly restorable since the lifetime of air plasma is within nanoseconds,[9] which means that once the driven laser is turned off, plasma ions and electrons recombine immediately and then the ambient air returns to its original state, allowing repeated reconfiguration of the plasma waveguides.

Yet for now, although the establishment of plasma arrays inside air is routine,[11-13] the underlying mechanism of THz field guidance along one single laser filament region is still not comprehensively understood, which further forms the obstacle for implementing interactions between double THz channels. For this reason, building blocks of the air-plasma THz computing circuit with basic functionalities (like calculus) remain absent and challenging. Fortunately, recently the epsilon-near-zero (ENZ, Re[$\varepsilon_r$]~0) effect of the air plasma filament in THz band was revealed,[14] which has not only interpreted the near-field strong confinement of THz amplitude during laser filamentation in air, but also opened the pathway to designing fundamental plasma devices for all-optical THz signal processing based on the free-space platform.

In this work, we firstly numerically built the two-filament array with a THz-ENZ zone generated inside this air-plasma structure. Then, we varied the air separation (in sub-millimeter scale) between the parallel plasma columns in order to optimize the spatial confinement and guidance of THz waves along the ENZ channel. Meanwhile, the THz output was fine adjusted with an 1/$f$-shape spectral response ($f$: THz frequency), which in principle of mathematics gives birth to a temporal integration of the broadband THz signal (in Section 2). One schematic diagram of this all-optical computing stage is displayed in **Figure 1**, inside which a single-cycle THz pulse ($E_{in}$) in time domain is integrated in form of $E_{out} = \int E_{in}dt$ via the designed dual-filament circuitry. At last, the above theoretical expectations have been experimentally confirmed in Section 3-5. Furthermore, this free-space laser-guiding-THz signal manipulation is promising to validate all-optical computing during remotely directional transmission of THz wave in air,[15] by which means the delivered THz information can be processed by remote control.

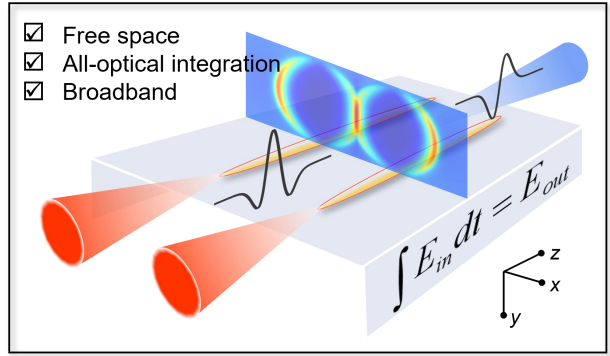

**Figure 1**. The concept of all-optical THz integration via the air-plasma platform, which consists of two identical parallel laser filaments separated by a sub-millimeter gap. The transverse figure in $x$-$y$ plane indicates the confined THz mode field mainly between the double filaments.

**2. The ENZ zone of air-plasma filaments with integration functionality**

In principle of mathematics as demonstrated in Supplementary Materials A, for an ideal time integration device, whose output temporal signal ($E_{out}$) is the first-order time integration of the input $E_{in}$, the ratio of spectral signals satisfies the 1/$f$ relation as

$$\frac{\mathcal{F}[E_{out}]}{\mathcal{F}[E_{in}]} \propto \frac{1}{f} \tag{1}$$

This mathematical relationship can be experimentally realized by near-field light confinement approaches.[3,16] Briefly, inside a near-field spatial region, the input lower-frequency component (with larger wavelength) was more tightly constrained with a stronger mode-field output, which resulted in the desired $1/f$ spectral response. In this direction, the ENZ effect is also well-known for its mode field confinement capability.[17,18] Besides, recently we found that the laser plasma filament could serve as an ENZ carrier of THz waves.[14] Thus, it is natural for us to consider the possibility of the plasma-based THz integration.

In order to investigate the ENZ region of the plasma filament, as well as interactions between two THz channels for potential all-optical integration, we carried out numerical simulations of the constrained THz mode within the air-plasma area. Firstly, in case of a single laser filament, it could be simplified as a plasma column with radial free-electron-density distribution ($N_e$) determined by a two-dimensional Gaussian function as

$$N_e = N_{e0} \exp(-\frac{x^2 + y^2}{b^2}) \tag{2}$$

where $N_{e0}$ denotes the central plasma density set at $1.0 \times 10^{17}$ cm$^{-3}$, and $b = 30$ μm is the half width at $1/e$ of the plasma column, whose full width at half maximum (FWHM) is about 50 μm. The radial distribution of $N_e$ is plotted in **Figure 2**a, according to which, the relative dielectric constant $\varepsilon_r$ is calculated by

$$\varepsilon_r = 1 - \frac{\omega_p^2}{\omega^2 - iv\omega} \tag{3}$$

where $\omega$ is the THz angular frequency and $\omega_p$ indicates the plasma frequency given by

$$\omega_p = \sqrt{\frac{e^2}{m_e \varepsilon_0} N_e} \tag{4}$$

where $e$ represents the electric charge, $m_e$ indicates the effective mass of the electron, and $\varepsilon_0$ is the permittivity in vacuum. $v \sim 1$ THz corresponds to the typical electron collision frequency inside the filament.[19] Figure 2b displays the radial distribution of Re[$\varepsilon_r$] inside the plasma filament, which increases towards the filament periphery and approaches the unity. And there is a positive-negative dielectric interface (Re[$\varepsilon_r$] = 0) near the filament periphery at approximately $r = 58$ μm (see inset).

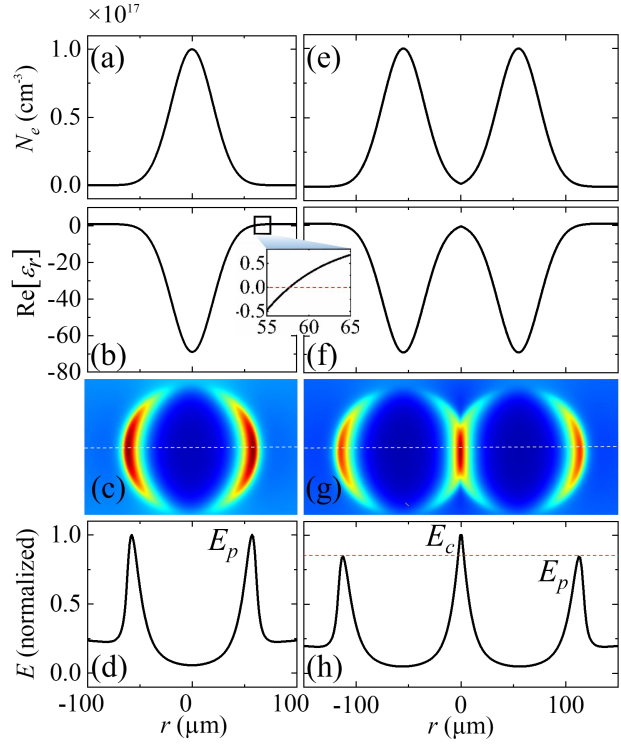

**Figure 2.** For a single plasma filament: the radial distribution of plasma density $N_e$ a) and the real part of dielectric constant $\varepsilon_r$ b), and the simulated electric field distribution of the confined THz mode at 0.3 THz c) with its radial profile along $y = 0$ d). e)-h) The corresponding results in case of the parallel double filaments array.

Based on this Re[$\varepsilon_r$] evolution, numerical simulations were carried out by the COMSOL software, and the simulated mode field at $f = 0.3$ THz is shown in Figure 2c. Then, the radial electric field distribution $E_r$ along $y = 0$ was further taken out and shown in Figure 2d. In view of Figure 2c and d, $E_r$ has been mainly confined close to the filament periphery where Re[$\varepsilon_r$] = 0, instead of the filament center. This result agrees with the 1DND analytical model as indicated in Supplementary Materials B.

Next, in order to enhance the THz wave confinement, similar simulations were also performed for the case of parallel plasma filaments separated by a distance of $2d \sim 100$ μm. The radial $N_e$ distribution of the double filaments is determined by

$$N_e = N_{e0} \exp\left\{-\frac{[x - d \cdot \operatorname{sgn}(x)]^2 + y^2}{b^2}\right\} \tag{5}$$

where the filament centers are along $y = 0$ and sgn($x$) denotes the sign of $x$.

Corresponding distributions of $N_e$ and Re[$\varepsilon_r$] are plotted in Figure 2e and f, respectively. Similar with the single filament case, there is Re[$\varepsilon_r$] = 0 at the filament periphery. Moreover, in the gap area of the double filaments, Re[$\varepsilon_r$] also approached zero, together with dramatic decreasing towards the two filaments' centers. This Re[$\varepsilon_r$] distribution has led to such a simulated result as shown in Figure 2g. A tighter confinement of the THz mode field has been induced by the synthesis of two THz-ENZ channels inside the filaments gap. This point is further highlighted in Figure 2h with $E_r$ along $y$ = 0, where the central THz electric field ($E_c$) of the filaments' gap mode is higher than the peripheral one ($E_p$) formed by a single filament.

To optimize this laser-guiding-THz interactions, the air separation (2$d$) between the plasma columns was varied from 90 μm to 125 μm during simulations (Supplementary Materials C). Meanwhile, $E_c$ and $E_p$ were extracted and shown in **Figure 3**a, inside which $E_c$ peaked at around 2$d$ = 110 μm while $E_p$ slightly decreased with respect to 2$d$. Furthermore, we also calculated $E_c/E_p$ as shown in Figure 3b, which was used to evaluate the enhancement factor of spatial THz confinement by the double filaments array compared with the single one. One can see that the range for $E_c/E_p$ above the unity (dotted red line) appeared around 2$d$ = 110 μm. Therefore, this 2$d$ result was adopted for our following simulations and experiments.

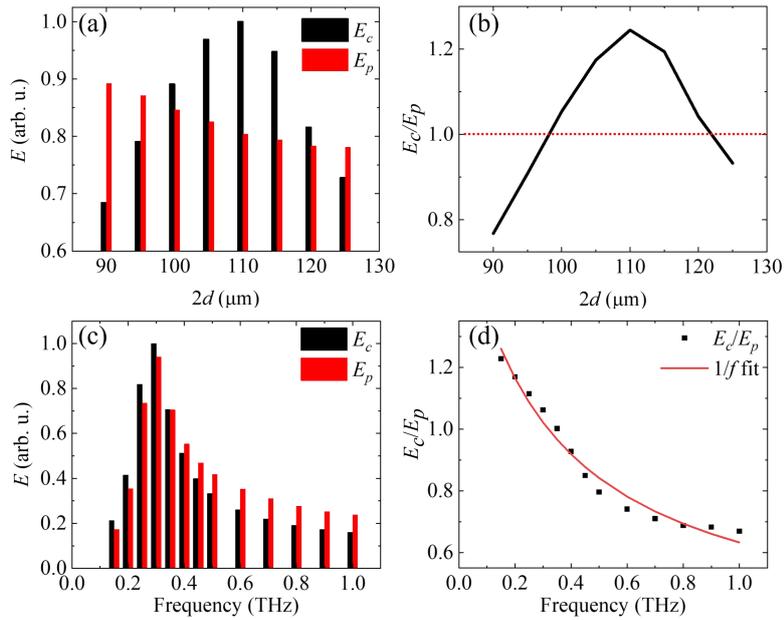

**Figure 3**. The central THz electric field ($E_c$, black bars) and the peripheral one ($E_p$, red bars) as a function of a) the distance 2*d* between two filaments and c) the THz frequency. The calculated $E_c/E_p$ as a function of b) 2*d* or d) THz frequency, and the corresponding 1/*f* fitting curve (red line).

The next concern is the frequency-dependent of $E_c/E_p$. This issue was clarified by repeating simulations within 0.1 to 1 THz at 2*d* = 110 μm (Supplementary Materials D). The resultant $E_c$ and $E_p$ are shown in Figure 3c, both of which firstly increase and subsequently decrease. It can also be seen that $E_p$ is less than $E_c$ in low THz band, but increases to exceed $E_c$ after 0.4 THz. Finally, the calculated $E_c/E_p$ are shown in Figure 3d as black squares, which indicate a decreasing spectral response as a function of THz frequency as expected. Remarkably, this result can be fitted by a 1/*f* curve, i.e.,

$$\left|\frac{E_c(f)}{E_p(f)}\right| \propto \frac{1}{f} \tag{6}$$

as shown as the red line in Figure 3d. Since Equation 6 meets the mathematical expectation of Equation 1, it is promising to perform time integration of THz signals by the ENZ-based filaments array. It is noticed that very recently, ENZ metamaterials with designed transmission functions were used to perform calculus on input signals,[20] while its electronic circuits and mechanism are still applied for the single frequency point, different from our broadband method in free space.

**3. Time-domain THz integration inside the dual-filament array**

To experimentally confirm the above points, a laser-pumping THz time-domain spectroscopy system[21] was established for both generation of the filament array and acquisition of the output THz signal. Specifically, a repetition rate of 1 kHz, central wavelength of 800 nm and duration of 100 fs Ti: sapphire laser pulse with energy of 2 mJ/pulse was split into two paths as the pump beam (~1.8 mJ/pulse) and the probe, respectively. The former was focused by a lens with focal length of 30 cm. Before and after the lens, a semicircular phase plate[11] and a 0.1-mm-thick Type I *β*-barium borate (BBO) crystal for frequency doubling from *ω* to 2*ω*, were inserted into the pump path. Thus, around the geometric focus of the lens, a 12-mm-length parallel double filaments were generated with separation distance of about 115 μm (see Supplementary Materials E).

It is worth noting that, besides the filament array playing the role of THz-ENZ waveguide, each of the plasma filaments radiated THz waves by itself, which can be retrieved by simply blocking one half of the phase plate and passing the pumping laser through the other half. Under this condition, two filaments were created alternately, and the emitted THz waveforms ($E_1$ and $E_2$) were recorded, respectively, by the probe beam in a standard EOS (electric-optic sampling) setup.[21] As shown in the inset of **Figure 4**a, $E_1$ and $E_2$ were both single-cycle temporal THz pulses with similar profiles. Moreover, the building time of these THz pulses (around a few picoseconds) were far behind the plasma column, which was produced within the same temporal scale of the pumping laser pulse duration, [22] i.e., ~100 fs. Thus, the plasma array can be treated as a static THz waveguide, along which $E_1$ and $E_2$ were both confined inside the central ENZ zone as simulated in Figure 2. Arising from this actuality, the output THz signal ($E_{out}$) after the double-filament array is expected to be in form of $E_{out} = \int E_{in} dt$, where $E_{in}$ is contributed by $E_1$ and $E_2$. Since $E_1$ and $E_2$ are in similar profiles, we adopted their average waveform as $E_{in}$ of this all-optical system, as displayed in Figure 4a as the black line. On the other hand, $E_{out}$ was detected with the filaments array as shown in Figure 4a as the red line. It is also noticed that waveforms in Fig. 4 are mostly normalized for analyses. This is reasonable because for the temporal signal and its calculus signal, quantitative comparisons are meaningless.

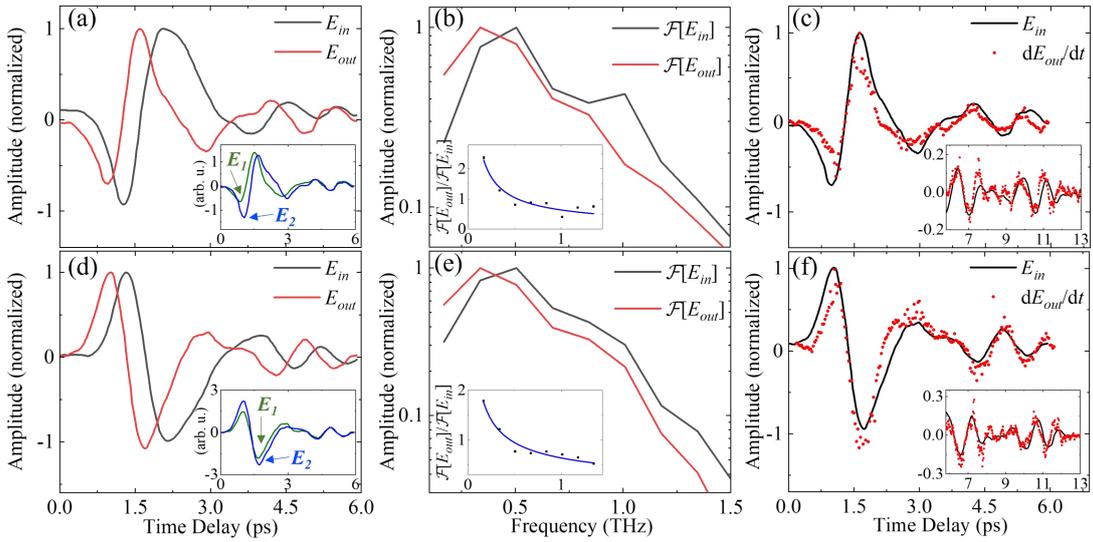

**Figure 4**. a) THz waveforms from the double filaments array (the red line, as $E_{out}$) and the average signal (the black line, as $E_{in}$) achieved from $E_1$ and $E_2$ of each one single filament (inset). b) THz spectra of $E_{in}$ ($\mathcal{F}[E_{in}]$, the black line) and $E_{out}$ ($\mathcal{F}[E_{out}]$, the red line) obtained by Fourier transforms on signals in a. Inset: $\mathcal{F}[E_{out}]/\mathcal{F}[E_{in}]$

(black squares) and its 1/*f*-fitted curve (the blue line). c) $E_{in}$ (the black line) and the first-order time derivative of $E_{out}$ (i.e., d$E_{out}$/d*t*, the red circles). Inset: oscillations of the THz waveform after the main pulse in repeated tests. d)-f) THz signals recorded in case of the BBO being moved towards the laser focus by a $\omega$-$2\omega$ dephasing length.

The normalized spectra of $E_{in}$ and $E_{out}$ via Fourier transforms are presented in Figure 4b, which shows a considerable low-frequency enhancement and high-frequency attenuation of $\mathcal{F}[E_{out}]$ with respect to $\mathcal{F}[E_{in}]$. The inset of Figure 4b depicts the ratio between $\mathcal{F}[E_{out}]$ and $\mathcal{F}[E_{in}]$, which unveils a basic match with the 1/*f* response associated to a time integrator as defined in Equation 1 and 6. To confirm this integration functionality, we calculated the first-order time derivative of $E_{out}$ (i.e., d$E_{out}$/d*t*) and compared it with the $E_{in}$ waveform, rather than integrating $E_{in}$ to match $E_{out}$. This strategy has the advantage of minimizing the effects of numerical artefacts, which could lead to alterations of the THz pulse shape (see Supplementary Materials F). The calculated d$E_{out}$/d*t* and $E_{in}$ waveforms are shown in Figure 4c with an agreement on the pulse profile. The inset of Figure 4c further proved this point with the signal oscillations after the main THz pulse in repetitive tests. Therefore, the capability of time-domain integrations has been demonstrated by the parallel air-filament array in THz band.

**4. Flexible conversion between THz integration and superposition calculation models**

In this section, we presented that simple changes of the BBO-to-focus distance (BFD) were able to switch the current THz integration model into the coherent superposition scheme,[23, 24] revealing the versatility of our all-optical system. To quantificationally distinguish these two models, the root mean squared error $\mathrm{RMSE}(x_i, X_i) = \sqrt{\dfrac{1}{n}\sum_{i=1}^{n}(x_i - X_i)^2}$ was used to estimate the matching between $E_{in}$ (average waveform of $E_1$ and $E_2$, and normalized) and d$E_{out}$/d*t* (normalized) — as RMSE$_1$, or between $E_{in}$ (= $E_1$ + $E_2$, non-normalized) and $E_{out}$ (non-normalized) — as RMSE$_2$. Here, *n* is the number of THz waveform samples, and $x_i$ is each sample value of $E_{in}$, and $X_i$ is that of d$E_{out}$/d*t* (or $E_{out}$).

By repeating the experiments with different BBO locations, RMSE$_1$ and RMSE$_2$ are calculated and shown in **Figure 5**a as black and red bars, respectively, as a function of BFD. Specifically, RMSE$_1$ achieved minimum at BFD = 20.5 cm, which was previously set in Section 3, while $RMSE_2$ peaked at this BBO position. Thus, here the THz integration model dominated. When there was a deviation from this BFD value, $RMSE_1$ rapidly increased with the decreasing of $RMSE_2$. Hence on this occasion, the original integration model was converted into the coherent addition pattern, which took the leading role. Overall, Figure 5a clearly illuminates the flexible switchover between two all-optical THz computing models merely by moving the BBO location of the air-plasma platform.

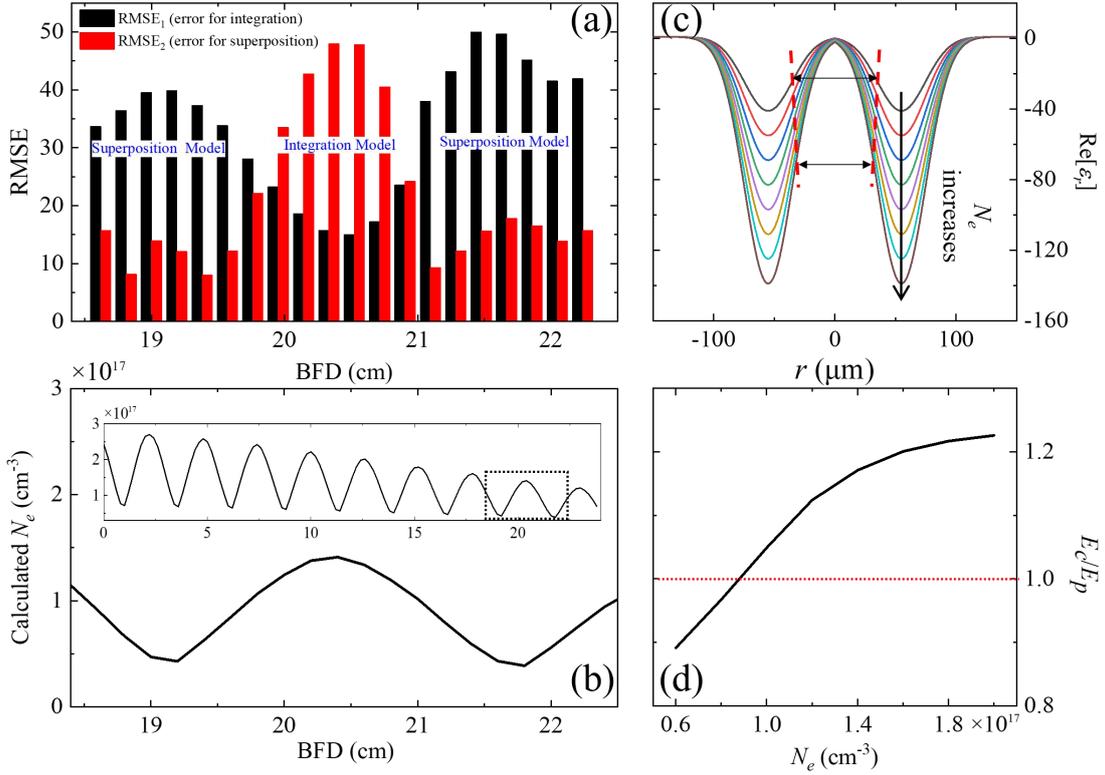

**Figure 5**. While BFD (BBO-to-focus distance) was varied: a) the calculated $RMSE_1$ for $E_{in}$ versus $dE_{out}/dt$ (black bars) and $RMSE_2$ for $E_{in}$ versus $E_{out}$ (red bars), and b) the calculated $N_e$. Inset: the overall $N_e$ as a function of BFD given by the photocurrent model. c) Re[$\varepsilon_r$] along $y = 0$ at different $N_e$. From top to bottom, $N_e$ varied from $0.6 \times 10^{17}$ cm$^{-3}$ to $2.0 \times 10^{17}$ cm$^{-3}$. d) The simulated $E_c/E_p$ as a function of $N_e$ according to the COMSOL model established upon c.

For further uncovering the underlying mechanism of this conversion, $N_e$ as a function of BFD has been calculated according to the static tunneling ionization method during photocurrent model calculations,[22, 25] and is shown in the inset of Figure 5b (see also Supplementary Materials G).

The detailed information inside the dashed box is enlarged and shown in Figure 5b. By visiting the correlation between Figure 5a and b, we noticed that $RMSE_1$ is inversely proportional to the $N_e$ value. In other words, the proposed THz integration effect might be strengthened by the increasing $N_e$. Reminding ourselves that at the very beginning, $N_e$ affected the $\varepsilon_r$ value via Equation 2-4, which then resulted in the ENZ zone ($Re[\varepsilon_r] = 0$) for the THz confinement. Hence, one can expect a potential correlation between $N_e$ and the integration effect.

In order to verify this prediction, we re-calculated the radial $Re[\varepsilon_r]$ with a varying $N_e$ ($0.6 \times 10^{17}$ cm$^{-3}$ ~ $2.0 \times 10^{17}$ cm$^{-3}$) at $f$ = 0.3 THz and $2d$ = 110 μm by Equation 3-5, and showed the results in Figure 5c. It can be observed that two $Re[\varepsilon_r]$ dips became deeper with the increasing $N_e$, and at the same time, the FWHM of the central confinement gap (at around $r$ = 0 where $Re[\varepsilon_r]$ approached zero) became narrower as highlighted by the red dashed lines. Based on these $Re[\varepsilon_r]$ profiles in Figure 5c, COMSOL simulation outcomes of $E_c/E_p$ are shown in Figure 5d, which increased with the increasing $N_e$ (Supplementary Materials H). Since this enhanced $E_c/E_p$ represented a tighter THz confinement inside the filaments' gap, as well as a stronger integration effect, thus, it can be concluded that the THz integration became more valid with the increasing $N_e$ in process of varying the BBO position. This is what actually can be observed in Figure 5a and b.

Furthermore, we noticed that $E_c/E_p$ started to be above 1 at about $N_e = 0.9 \times 10^{17}$ cm$^{-3}$, and tended to be saturated at $N_e = 2.0 \times 10^{17}$ cm$^{-3}$ as shown in Figure 5d. This phenomenon hints that the spatial confinement of THz energy between the double plasma filaments, as well as the following THz integration effect, can be significantly changed when $N_e$ just varied around twice. While the calculated $N_e$ maximum (~$1.4 \times 10^{17}$ cm$^{-3}$) in Figure 5b was more than three times larger than the minimum (~$0.4 \times 10^{17}$ cm$^{-3}$). Therefore, this $N_e$ variation range is able to give rise to the appearing or vanishing of the synthetic THz mode in gap (integration phenomenon). And when $N_e$ was weak to support this gap mode, the two filaments had no more interactions between each other, so the integration effect would be overtaken by the alternative THz superposition model from two independent filaments, as can be seen in Figure 5a.

In consideration of the above analyses, we additionally checked the THz signals at another $N_e$

maximum with the BBO position moved towards the laser focus by a $\omega$-$2\omega$ dephasing length[26] ($l_d$ ~ 27 mm in our experiment). And the time-domain THz integration was again observed as indicated in Figure 4d-f, well supporting our theory. Comparing the THz waveforms between Figure 4c and f, the polarities of these two THz electric fields were completely reversed, which was induced by the change of BFD by a dephasing length as expected in Ref. [27].

**5. Time-domain THz integration inside the four-filament array**

One advantage of the proposed all-optical computing stage is the rapid reversibility of the air plasma, which can be either built up within the same temporal scale of the pumping laser pulse duration, i.e., a few tens of femtoseconds, or erased in nanoseconds (i.e., the lifetime of plasma) as long as the driven laser is blocked. This could be much quicker than the relaxation time of the reported LGL materials (e.g., gel[6]). Thus, theoretically in our case, the reconfiguration of the system is only limited by the consuming time of common experimental operations, e.g., moving the BBO crystal to proper positions (Section 4) or changing the phase plate type (as follows).

In this section, the proposed time-domain THz integration effect has been explored by using a parallel four filaments array, which was produced by a quarter-circle phase plate (QCPP).[11] Similar procedures with the previous double-filament experiment were performed in this four filaments case. For example, this time, $E_{in}$ was the average of $E_1$, $E_2$, $E_3$ and $E_4$, which were the THz pulses radiated from each of the four filaments by blocking one quarter of the phase plate. And $E_{out}$ was the THz signal from the whole filament array. The final data are shown in **Figure 6**, from which one can see all the same characteristics aforementioned in Figure 4, including I the considerable enhancement of the low-frequency components of $E_{out}$ and (ii) the 1/$f$ tendency of $\mathcal{F}[E_{out}]/\mathcal{F}[E_{in}]$ as shown in Figure 6a and c, and (iii) the satisfying match between $E_{in}$ and d$E_{out}$/d$t$, namely, $E_{out}$ is the integration of $E_{in}$ in time domain as shown in Figure 6b and d, and (iv) the inverse symmetry of THz waveforms given by two BFDs separated by a dephasing length as shown in Figure 6b and d.

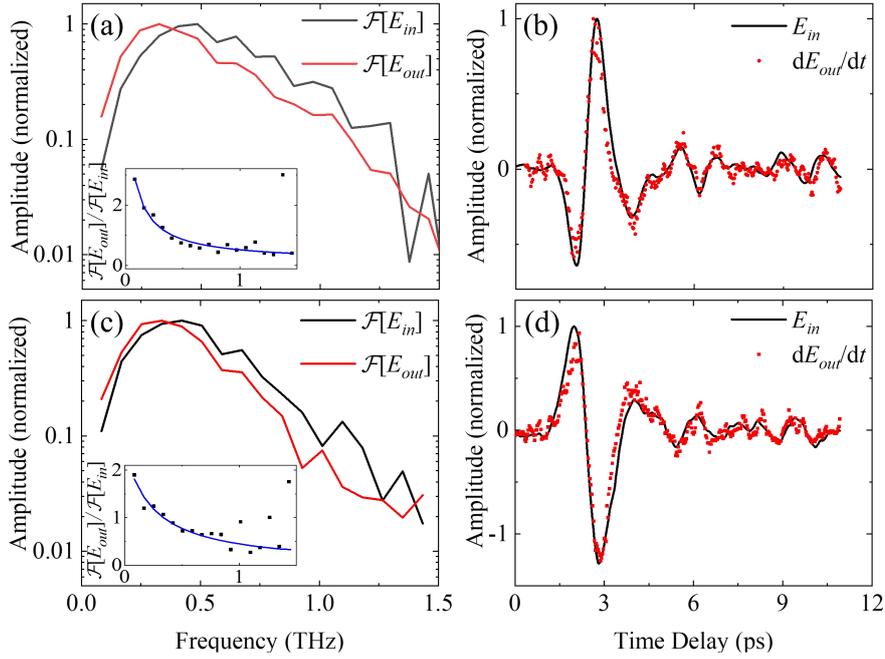

**Figure 6**. For the four filaments array: a) THz spectra of $E_{in}$ ($\mathcal{F}[E_{in}]$, the black line) and $E_{out}$ ($\mathcal{F}[E_{out}]$, the red line). Inset: $\mathcal{F}[E_{out}]/\mathcal{F}[E_{in}]$ (black squares) and its $1/f$-fitted curve (the blue line). b) $E_{in}$ (the black line) and the first-order time derivative of $E_{out}$ (i.e., $dE_{out}/dt$, the red circles). c)-d) THz signals recorded in case of the BBO being moved towards the laser focus by a $\omega$-$2\omega$ dephasing length.

Compared with the double filaments array, the four filaments seemed to be a better time integrator, since the calculated RMSE between $E_{in}$ and $dE_{out}/dt$ in Figure 6b is only 8%. This can be attributed to the stronger THz confinement contributed by more ENZ zones between each two of the four filaments (Supplementary Materials I).

## 6. Conclusion and outlook

In summary, in this work we developed a non-static platform for LGL integration of broadband THz waves solely based on the air plasma. Briefly, this all-optical stage was dynamically constructed by laser-ionization-induced transient plasma waveguide circuits in flexible conversion between two- and four-filament array due to the reversibility of air. Then, the THz signal was spatially confined along the ENZ area of the plasma structure, resulting in a spectral response in $1/f$ shape which accounted for the final integration phenomenon. The presented free-space LGL scheme for optical signal processing has been realized by interactions between light beams without assistance from conventional solid waveguide circuits, thus paving the way towards new approaches of all-optical computing without complex hardwares and algorithms. In future developments, our prototype setup holds the potential to become a building block for constructions of a wide range of analog systems for signal transmission, processing and computing applications based on THz waves.

Moreover, the suggested all-optical method can not only serve as a tabletop computing chip with sub-millimeter-scale information channel, but also can be applied during remotely directional transmission of THz waves in air, e.g., cascaded with the free-space photonic crystal (FPC).[15] In this case, the unwanted weak THz differentiation effect of the FPC waveguide (Supplementary Materials J) could be counteracted by the following integration given by the double-filament stage. Thus, a THz signal far away after long-distance guided travel can be recovered to its original shape by remotely controlling the all-optical-filaments setup combined by the FPC and two-filament array. If further mixed with the BFD-tuned superposition model (in Section 4) and the subtraction pattern by crossing two filaments,[28] it is promising to realize an air-plasma-based all-optical hybrid computation station in free space.

**Acknowledgement**

This work was supported in part by the Youth Sci-Tech "Qimingxing" (22QC1400300) Program of Shanghai, National Natural Science Foundation of China (61988102), 111 Project (D18014), International Joint Lab Program supported by Science and Technology Commission Shanghai Municipality (17590750300), Key project supported by Science and Technology Commission Shanghai Municipality (YDZX20193100004960), and the General Administration of Customs Project (2019hk006, 2020hk251).**References**

# Supplementary Materials
## of
## Air-plasma-based all-optical temporal integration of broadband terahertz pulses

Jia-Yu Zhao

**A. Transfer functions of an integrator/differentiator for broadband temporal pulses**

The Fourier transform of a temporal pulse $f(\tau)$ can be expressed as

$$\mathcal{F}[f(\tau)] = \int_{-\infty}^{\infty} f(\tau)e^{-j\omega\tau}d\tau = F(\omega) \tag{s1}$$

And the integration of $f(\tau)$ is

$$g(t) = \int_{-\infty}^{t} f(\tau)d\tau \tag{s2}$$

Thus, the Fourier transform of $g(t)$ can be calculated by

$$\mathcal{F}[g(t)] = \int_{-\infty}^{\infty} g(t)e^{-j\omega t}dt = \int_{-\infty}^{\infty}[\int_{-\infty}^{t} f(\tau)d\tau]e^{-j\omega t}dt$$

$$= \int_{-\infty}^{\infty}[\int_{-\infty}^{\infty} f(\tau)step(t-\tau)d\tau]e^{-j\omega t}dt \tag{s3}$$

$$= \int_{-\infty}^{\infty} f(\tau)e^{-j\omega\tau}d\tau \cdot \int_{-\infty}^{\infty} step(t-\tau)e^{-j\omega(t-\tau)}d(t-\tau)$$

where $step(t-\tau)$ is the step function and its Fourier transform is $[\pi\delta(\omega) + 1/(j\omega)]$. Hence, Equation (s3) can be further transformed into

$$\mathcal{F}[g(t)] = \pi\delta(\omega)F(0) + \frac{1}{j\omega}F(\omega) \tag{s4}$$

At last, the transfer function of a first-order time integrator is

$$H(\omega) = \mathcal{F}[g(t)] / \mathcal{F}[f(\tau)] \propto \frac{1}{j\omega}F(\omega) / F(\omega) = \frac{1}{j\omega} \tag{s5}$$

which is schematically shown in Figure s1a.

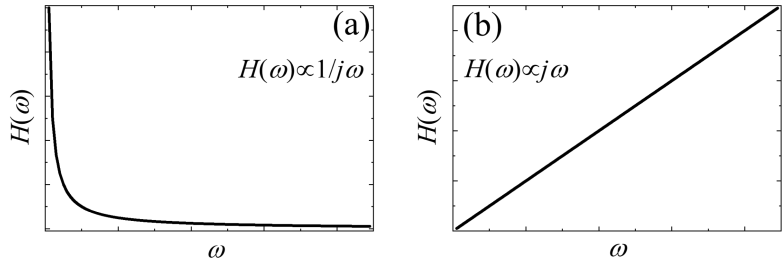

**Figure s1.** The transfer function of a) a first-order time integrator and b) a first-order time differentiator.

On the other hand, the inverse Fourier transform of $f(\tau)$ can be expressed as

$$f(\tau) = \frac{1}{2\pi} \int_{-\infty}^{\infty} F(\omega) e^{j\omega\tau} d\omega \qquad (s6)$$

If we differentiate Equation (s6) and could get

$$g(\tau) = \frac{df(\tau)}{d\tau} = \frac{1}{2\pi} \int_{-\infty}^{\infty} [j\omega F(\omega)] e^{j\omega\tau} d\omega \qquad (s7)$$

From Equation (s7), we can easily observe

$$\mathcal{F}[g(\tau)] = j\omega F(\omega) \qquad (s8)$$

As a consequence, the transfer function of a first-order time differentiator is

$$H(\omega) = \mathcal{F}[g(\tau)] / \mathcal{F}[f(\tau)] = j\omega F(\omega) / F(\omega) = j\omega \qquad (s9)$$

This transfer function of a first-order time differentiator is shown in Figure s1b.

**B. Verification of the numerical simulations by an analytical model**

To verify the COMSOL simulations in the main text, we employed an analytical model (1-D negative dielectric waveguide, 1DND) [s1] to reproduce the THz mode field with same parameters. As shown in Figure s2, the 1DND-calculated result obtains radial modal maximums at the filament periphery around $r = \pm 56$ μm, where $\mathrm{Re}[\varepsilon_r] = 0$, thus sharing similar properties with the numerical outcomes in the main text [Figure 2d].

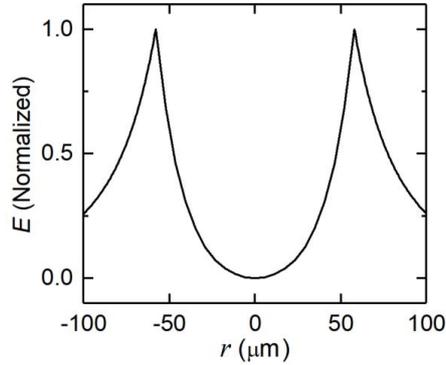

**Figure s2.** The radial THz mode distribution given by the 1DND model.

**C. Determining the proper separation distance between two filaments**

In order to determine the separation distance (2$d$) between two filaments for a tight THz confinement, we simulated the THz mode with 2$d$ increasing from 90 μm to 125 μm as shown in Figure s3. It can be observed that the central electric field ($E_c$) firstly increased and reached maximum at 2$d$ = 110 μm but then decreased, while the peripheral one ($E_p$) always decreased during this simulation. The quantitative $E_c$ and $E_p$ values are shown in Figure 3a of the main text, and 2$d$ = 110 μm has been selected in the subsequent simulations and experiments for its strongest THz confinement between two filaments.

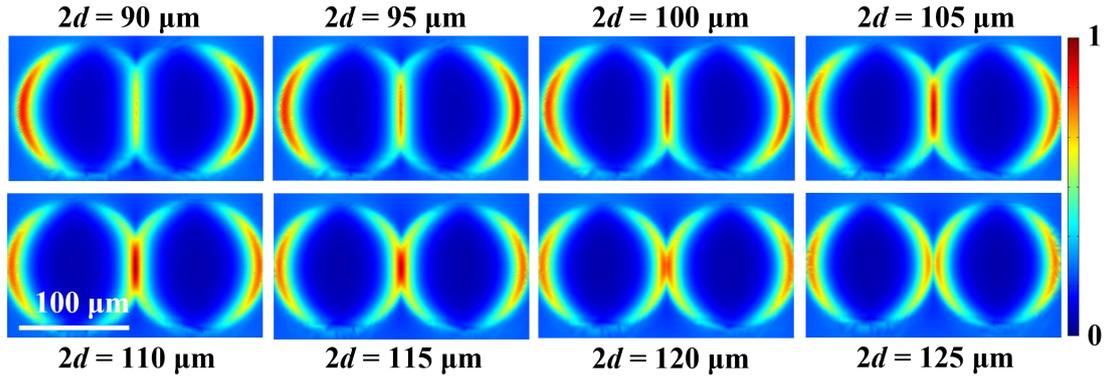

**Figure s3**. The simulated THz mode at 0.3 THz as $2d$ varied between 90 μm and 125 μm.

### D. The simulated THz mode at different frequencies

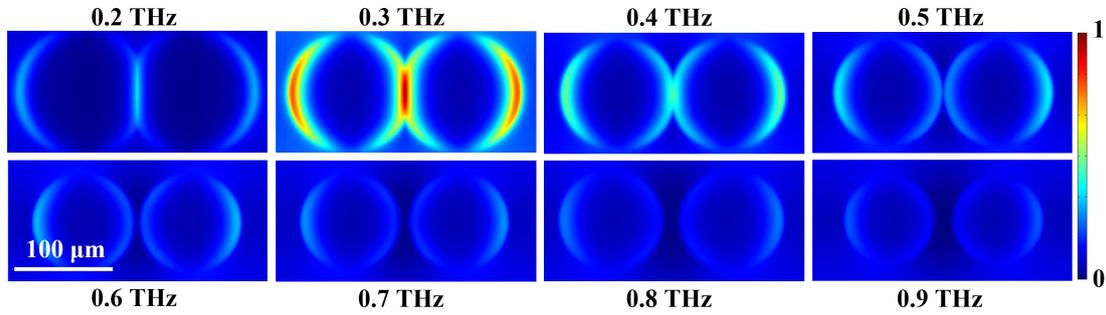

**Figure s4**. Parts of the simulated THz modes ranging from 0.1~1 THz at $2d$ = 110 μm.

### E. The experimental realization of double filaments with the selected separation distance

In Section C, the simulated $2d$ = 110 μm was decided for following works. Next, in order to experimentally realize this issue, a phase plate, which is divided into two parts with π phase lag at 800 nm [s2], was used by us to create the double filaments array. Moreover, according to the theory in Ref. [s2], the separation distance between filaments is mainly determined by the focal length of focusing lens, which should be shorter than 50 cm if $2d$ = 110 μm is the target. In our tests, the focal length of 30 cm was finally employed, and the detected $2d$ was about 115 μm. This can be seen from the resultant two filaments, which were captured by both the thermal paper and the side image methods as shown in Figure s5.

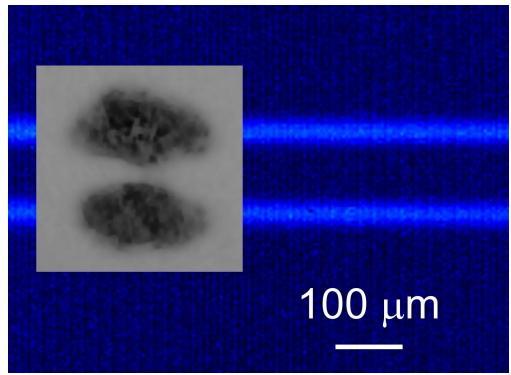

**Figure s5**. The side image of the generated two filaments. Inset: the burning pattern on the thermal paper by inserting it transversely into the two-filament array.

## F. Comparisons between two methods of data processing

In Section 3 of the main text, we mentioned that the strategy of comparing $E_{in}$ and $dE_{out}/dt$ "has the advantage of minimizing the effects of numerical artefacts, which could lead to alterations of the THz pulse shape". Here, we used the following Figure s6 and s7 to express this point. For the double filaments array, Figure s6a and c are given by $E_{in}$ and $dE_{out}/dt$, same with Figure 4c and f in the main text. On the other hand, Figure s6b and d are given by $\int E_{in}\, dt$ and $E_{out}$. It is evident that the latter method led to alterations of the waveform shape due to the superposition of a spurious offset value during $\int E_{in}\, dt$. This phenomenon is more obvious in Figure s7 for the four-filament array.

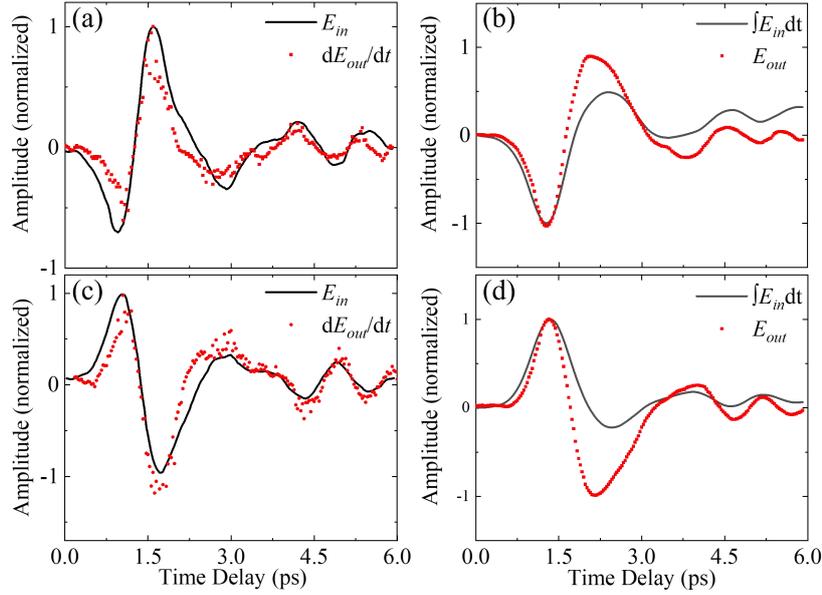

**Figure s6**. The double-filament array: a) c) For the data processing method of differentiating the output signal as in the main text; b) d) For the data processing method of integrating the input pulse.

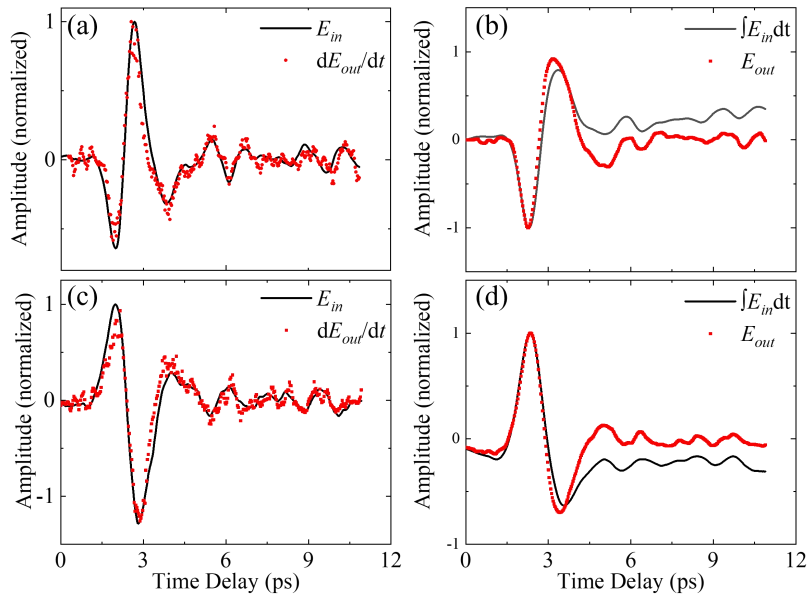

**Figure s7**. Corresponding cases for the four-filament array.

## G. Verification of the photo-current modal calculations

To validate our calculations via the photo-current model in this work, we additionally calculated the THz peak-to-peak (P-P) amplitude with this model, and also measured the experimental THz P-P amplitude. As plotted in Figure s8, the theoretical and experimental values agreed well with each other. More importantly, our results shared similar characteristics with the highly-cited Ref. [s3]:

(1) It is clear that the THz P-P amplitude exhibited a sine-shaped behavior as a function of BFD, which is due to the variation of relative phase difference between the fundamental laser ($\omega$) and its second harmonic ($2\omega$) carriers during propagation in air.

(2) The THz P-P amplitude experienced an overall decline as BFD increased, which is highlighted by the dashed line in Figure s8. This is resulted from the increasing mismatching between the $\omega$ and $2\omega$ pulse envelopes.

(3) The THz P-P amplitude appeared to be minimum around BFD = 0. This is a typical consequence predicted by the photo-current model.

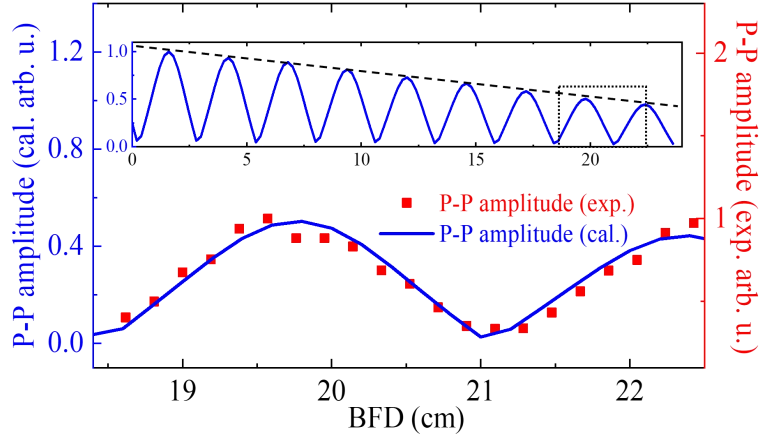

**Figure s8**. While BFD (BBO-to-focus distance) being varied: the experimental peak-to-peak (P-P) amplitude of THz pulses (squares), and the calculated THz P-P amplitude (solid line) via the photo-current model. Inset: the full view of the modal results.

## H. The simulated THz mode at different $N_e$

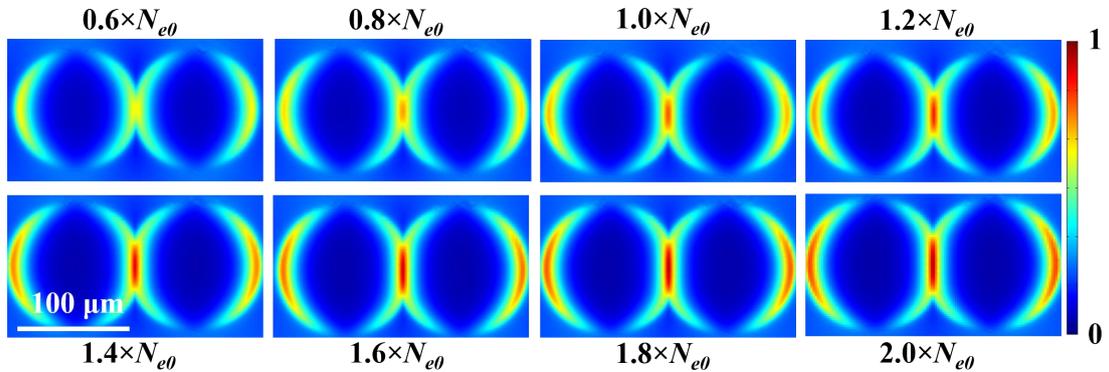

**Figure s9**. The simulated THz mode as $N_e$ increased from $0.6 \times N_{e0}$ to $2.0 \times N_{e0}$, where $N_{e0} = 1 \times 10^{17}$ cm$^{-3}$.

## I. The simulated THz mode within a four-filament array

We numerically simulated the THz confinement effect inside a four-filament array, and the corresponding field distribution is shown in Figure s10. Note that, here the $N_e$ distribution of each filament was set to be uniform for highlighting the confined THz field inside four ENZ gaps. Compared with the case of double filaments with only one gap area, this four-filament structure succeeded by its stronger THz confinement. This could be the reason for its better integration result as shown in Figure 6 of the main text.

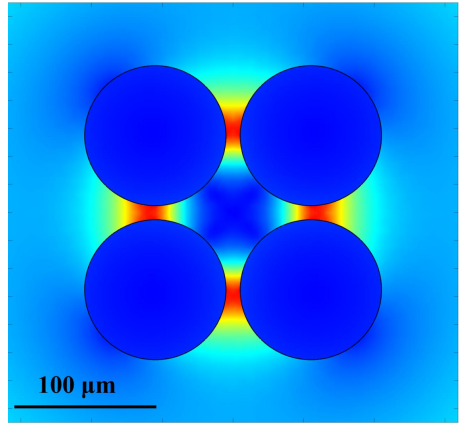

**Figure s10**. The simulated THz mode in the four filaments array

## J. The concept of free-space photonic crystal (FPC)

The free-space photonic crystal (FPC) waveguide has been previously proposed for remotely directional transmission of THz waves in Ref. [s4]. The structural features and the simulated THz mode are shown in Figure s11a and b, respectively. Moreover, its spectral transfer function increased with the increasing THz frequency as shown in Figure s11c. This implies that the FPC channel is roughly a candidate for THz differentiation devices as indicated by Figure s1b and Equation s9. In view of this issue, this FPC structure (with weak differentiation effect) could be connected with the suggested two/four-filament array in this work (with integration effect) for the purpose of remotely modulating the long-distance transmitted THz signal to its original shape.

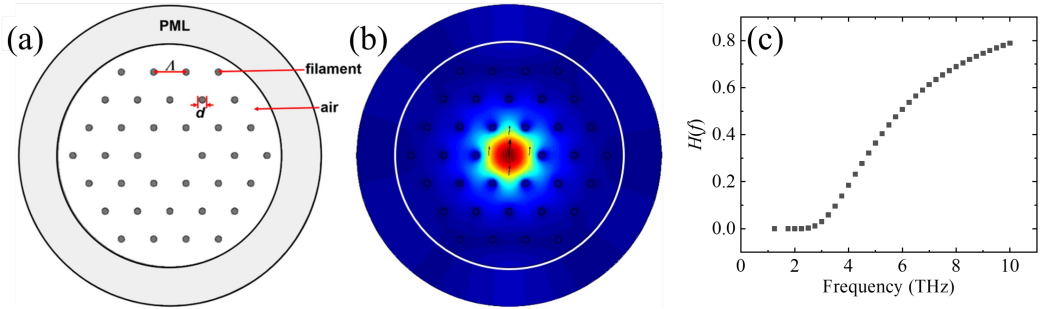

**Figure s11.** a) Structural parameters of the FPC array: $d$ = 100 μm and $\Lambda$ = 500 μm. b) The simulated THz mode profile at 6 THz and $N_e$ = $10^{17}$ cm$^{-3}$. c) The transfer function of the FPC waveguide.